\documentclass[review,3p,times]{elsarticle}

\usepackage{amsmath,hyperref}
\usepackage{lineno}

\usepackage{epstopdf}
\journal{Physical Review E}

\usepackage{graphicx}
\usepackage{dcolumn}
\usepackage{bm}
\usepackage{epsfig}
\usepackage{booktabs}
\usepackage{graphics}
\usepackage{amssymb}
\usepackage{amsmath}
\usepackage{array}
\usepackage{color}
\usepackage{booktabs}
\usepackage{multirow}
\usepackage{caption}
\usepackage{chngpage}
\usepackage{subcaption}
\usepackage{mathrsfs,gensymb,float}
\usepackage{algorithm, algorithmic}
\usepackage[T1]{fontenc}

\biboptions{numbers,sort&compress}



\begin{document}

\captionsetup[figure]{labelfont={bf},name={Fig.},labelsep=period}        

\begin{frontmatter}
	
	\title{Local volume-conserving lattice Boltzmann model for incompressible multiphase flows}
	\author[1]{Fang Xiong}
	\author[1]{Lei Wang\corref{mycorrespondingauthor}}
	\cortext[mycorrespondingauthor]{Corresponding author}
	\ead{wangleir1989@126.com}
	\author[2]{Xinyue Liu}
		
\address[1]{School of Mathematics and Physics, China University of Geosciences, Wuhan 430074, China}
\address[2]{Department of Mathematics, College of Science, National University of Defense Technology, Changsha, Hunan, 410073, China}

\begin{abstract}
The Cahn-Hilliard (C-H) equation, as a classical diffusion-interface method of phase-field, has been extensively employed for simulating two-phase fluid dynamics. However, it suffers from a key challenge in the simulation process, specifically the volume conservation of each phase cannot be guaranteed. To address this issue, in this paper, a modified C-H equation for two-phase flow modeling is first introduced, and the basic idea of this model lies in that it combines the profile correction method with the level-set approach, and thus, it effectively improves the deficiency of the classical C-H equation in terms of volume non-conservation of each phase. Based on this modified C-H equation, we further propose an accurate interface-capturing lattice Boltzmann (LB) model. After that, we perform a range of numerical simulations, including two stationary droplets immersed in the gas phase, single vortex, Rayleigh-Plateau fluid instability, and droplet deformation under a shear flow. These simulations illustrate that the proposed LB model has superior performance in maintaining local volume conservation and accurately capturing interfaces. More importantly, compared to the LB model derived from the classical C-H equation, it not only achieves more precise volume conservation for each phase but also provides a more consistent representation of the droplet's interface morphology more consistently, especially in dealing with small droplet problems. 

\end{abstract}

\begin{keyword}
Lattice Boltzmann method \sep Cahn-Hilliard equation \sep Local volume-conservation
\end{keyword}

\end{frontmatter}

\section{Introduction}
Two-phase flow phenomena are commonly observed in both natural settings and engineering applications, such as karst flow in volcanic eruptions \cite{Gonnermann2013Dynamics}, boiling \cite{Dhir1998Boiling}, the transportation of crude oil and gas-liquid separation \cite{Hart2014Areview}. However, the transport process of two-phase flow systems is complicated due to the complex changes in the interface, which include migration, deformation, breakup, and merging \cite{Thorsen2002Microfluidic}. Accurately describing and predicting the behavior of these two-phase interfaces remains a significant challenge. In recent decades, researchers have conducted extensive theoretical \cite{Cristini2004Theory, Elbar2025Weak} and experimental \cite{Steinke2004Anexperimental} to better understand the flow characteristics of two-phase systems. Nevertheless, theoretical analyses are often limited to simpler cases and struggle to address complex problems involving multi-field coupling, making it difficult to obtain accurate solutions \cite{Cristini2004Theory, Leshansky2012Obstructed}. While experimental methods are widely employed to capture the macroscopic dynamic behavior of interfaces, they typically lack the capability to detail the underlying flow processes, in addition to being constrained by high costs and lengthy durations for complex scenarios \cite{Link2004Geometrically, Jullien2009Droplet}. 

Driven by the swift progress in supercomputing capabilities and advanced computational techniques, numerical simulation has emerged as a highly effective method for modeling two-phase flow phenomena \cite{Yue2004Adiffuse, Hao2024Aninterfacial}. Numerical approaches to addressing two-phase flow problems generally fall into two primary categories: sharp interface methods \cite{Hirt1981Volume, Sussman1997Axisymmetric, Unverdi1992Afront} and diffuse interface methods \cite{Jacqmin1999Calculation, Jain2022Accurate}. Sharp interface methods consider the interface between two phases as a geometrically discontinuous surface with zero thickness. The fluid domain is split into two distinct single-phase regions, with each phase's hydrodynamic behavior described by its governing equations \cite{Zhang2019High-order}. Common sharp interface methods encompass the volume of fluid method \cite{Hirt1981Volume}, level-set method \cite{Sussman1997Axisymmetric, Sethian2003Level}, front tracking method \cite{Unverdi1992Afront}. The volume of fluid method \cite{Bonhomme2012Inertial}, while capable of handling complex changes in interfacial topology, often uses first-order reconstruction formats for interface \cite{Wang2019Abrief}, which can lead to discontinuities in the volume fraction function at the interface and potential numerical oscillations \cite{Wang2019Abrief}. In addition, the level-set method is effective for calculating curvature, normal vectors, and surface tension, but requires re-distancing when the interface topology change significantly, which may affect volume conservation \cite{Sussman1997Axisymmetric}. Moreover, although the front-tracking method \cite{Unverdi1992Afront} provides clear and high-order accurate interfaces, it becomes increasingly complicated when dealing with multi-phase flows \cite{Wang2019Abrief}. As opposed to sharp interface methods, diffuse interface methods substitute the sharp interface with a smooth transition zone of finite width \cite{Wang2019Abrief}. In this region, properties such as viscosity and density of the two phases vary gradually, eliminating the require for explicit interface tracking \cite{Wang2019Abrief}, which makes diffuse interface methods especially advantageous for studying two-phase flows with complex interfacial topology changes \cite{Wang2019Abrief}. In diffuse interface frameworks, the motion and evolution of the interface are described through an order parameter that adheres to the phase-field equations, among which the most widely used phase-field diffuse interface models is the so-called C-H equation \cite{Cahn1958Free}, and the distinct is that it ensures volume conservation in multiphase systems.

For the past few years, numerical methods based on diffuse interface models have gained considerable attention, among which, the LB method \cite{Krüger2017The, Wang2023Phase, Xiong2025Athermodynamically, Li2023Asimplified} has achieved great success due to its effectiveness in dealing with complex interfaces without the need for explicit tracking \cite{Wang2023Phase}. For two-phase flow LB models, two distinct types of distribution functions are essential: one for solving the hydrodynamic equations and another to address the phase-field equation \cite{Zhang2019High-order}. To accurately recover the phase-field equation, various LB models have been developed specifically for the C-H equation. Historically, the first endeavor to develop a LB model for the C-H equation can be traced to the groundbreaking research of He et al. \cite{He1999Alattice}, who describes the interface using an order parameter and tracked the interface with an index function. Subsequently, Lee et al. \cite{Lee2005Astable} proposes a similar LB model that uses density instead of the order parameter. They also introduce a numerical method to discretize the convection term, aiming to improve model stability. However, this approach encounters the same issue as that described by He et al. \cite{He1999Alattice}, it fails to fully align the LB model at the interface with the C-H equation \cite{Zheng2006Alattice}. To provide the correct macroscopic C-H equation, Zheng et al. \cite{Zheng2006Alattice, Zheng200Three}  introduced a novel LB model by incorporating a spatial difference term of the distribution function. Unfortunately, as noted by Fakhari and Rahimian \cite{Fakhari2010Phase-field}, this method is applicable only to a density-matched binary-fluid model that satisfies the Boussinesq approximation, which does not accurately represent actual two-phase systems. To overcome this restriction, Fakhari and Rahimian \cite{Fakhari2010Phase-field} proposed an improved LB model, that is highly sensitive to mobility values and introduces certain artifacts in the recovered C-H equation that affect the numerical accuracy \cite{Liang2014Phase}. After that, follow Zheng et al.'s \cite{Zheng2006Alattice} work, Zu and He \cite{Zu2013Phase} implemented a modification by replacing the distribution function with the spatial difference term of the equilibrium distribution function, aiming to enhance both accuracy and stability, but the model becomes prone to instability as the relaxation time nears 1.0 \cite{Liang2014Phase}. Different from Zu and He's work \cite{Zu2013Phase}, Liang et al. \cite{Liang2014Phase} incorporated a time-derivative term into the LB equation, eliminating the extraneous terms present in the C-H equation proposed by Lee et al. \cite{Lee2005Astable}. Recently, building on Liang et al.'s \cite{Liang2023Lattice} research, Ju et al. \cite{Ju2023Awell} redefined the C-H equilibrium distribution function by treating the convective term as a source term, achieving discrete force balance and reducing spurious velocities. 

As mentioned above, in the LB community, progress has been made in solving the classical C-H equation, however, there are still challenges to address \cite{Hao2024Aninterfacial}. As noticed by Hao et al. \cite{Hao2024Aninterfacial}, although the C-H equation naturally preserves mass conservation globally, the interaction between the convection term and the diffusion term can lead to thickening or thinning of the interface \cite{Hao2024Aninterfacial}. In addition, discrete errors in space and time at the interface can result in numerical diffusion, causing the fluid interface to deviate from its equilibrium position and leading to volume non-conservation in each phase \cite{Hao2024Aninterfacial}. These issues can significantly impact the precision of the interfacial surface tension as well as the depiction of fluid characteristics, thus creating uncertainty in simulation results, particularly in hydrodynamic modeling involving surface tension \cite{Hao2024Aninterfacial}. Further, Wang et al. \cite{Wang2015Amass} attempted to propose a multiphase LB flux solver to address the revised C-H equation, aiming to ensure mass conservation in each phase \cite{Wang2015Amass}. Specifically, the fundamental idea behind this method involves incorporating a mass correction term into the C-H equation, and the mass correction term is designed to compensate for any mass loss or gain that occurs due to numerical errors or modeled diffusion at each time step  \cite{Hao2024Aninterfacial}, which improves the mass non-conservation issue to some extent \cite{Wang2015Amass}. However, as pointed out by Mirjalili et al. \cite{Mirjalili2017Interface}, this correction makes the C-H equation no longer a gradient flow of energy generalization, and it may fail to effectively address the jump in the phase field function at the interface \cite{Hao2024Aninterfacial}.

Based on this background, this work intends to employ an improved two-phase interface profile preservation method that integrates the level-set method \cite{Chiu2011Aconservative, Chiu2019Acoupled, Jain2022Accurate} and a profile correction technique \cite{Li2016Aphase} to build a model for two-phase flow problems, which aims to reduce interface dispersion, enhance modeling accuracy, and minimize volume conservation errors in each phase. Subsequently, a generalized LB model is proposed to handle two-phase flow simulations. The structure of this work is outlined as follows. In Sect. 2, we will give the two-phase diffuse interface fluid model. In Sect. 3, the corresponding LB model is developed. In Sect. 4, some numerical simulations are conducted. At last, a concise summary is presented in Sect. 5.

\section{Two-phase diffuse interface fluid model}
Within the context of phase-field theory, the C-H equation is obtained through the minimization of the Landau free energy generalization, which is given by \cite{Cahn1958Free, Kendon2001Inertial}:
\begin{equation}
	{{\mathcal{F}}_\phi } = \int_\Omega  {\left( {{\psi_0} + \frac{\kappa }{2}{{\left| {\nabla \phi } \right|}^2}} \right)} d\Omega,
	\label{eq1}
\end{equation}
where $\Omega$ is the fluid domain. The bulk free-energy density $\psi_0$ satisfies a double-well potential of the form ${\psi_0} = \beta {\phi ^2}{\left( {\phi  - 1} \right)^2}$ \cite{Cahn1958Free, Jacqmin1999Calculation}, the formulation defines the order parameter $\phi$, with a value of 0 corresponding to the lighter fluid and a value of 1.0 representing the heavier fluid. $\kappa {{\left| {\nabla \phi } \right|}^2}/2$ denotes the additional interfacial energy at the fluid-fluid interface \cite{Jacqmin1999Calculation}. The parameters $\beta$ and $\kappa$ are coefficients tied to the interface thickness $W$ and interface surface tension $\gamma$, with the relationships $\beta=12\gamma/W$ and $\kappa=3\gamma\mathrm{W}/2$ \cite{Liang2014Phase}. The chemical potential is established by computing the variational derivative of free energy ${\mathcal{F}}_\phi$ concerning $\phi$ as
\begin{equation}
	\hbar = \frac{{\delta {\mathcal{F}}_\phi}}{{\delta \phi}} = 4 \beta\phi(\phi-1)(\phi-0.5)-\kappa\nabla^{2}\phi.
	\label{eq2}
\end{equation}
Therefore, the evolution of the order parameter $\phi$ is governed by the C-H equation as follow
\begin{equation}
	\frac{\partial\phi}{\partial t}+\nabla\cdot(\phi \mathbf{u})=\nabla\cdot(M\nabla\hbar),
	\label{eq3}
\end{equation}
in which, $t$, $M$ and $\mathbf{u}$ denote time, mobility and velocity, respectively. When the interface reaches equilibrium, i.e. $\hbar=0$, we get the equilibrium profile of the order parameter $\phi$ conforms to the hyperbolic tangent function \cite{Zhang2009Numerical}
\begin{equation}
	\phi=0.5[1+ \rm{tanh}(\frac{2s}{W})],
	\label{eq4}
\end{equation}
in which $s$ is the coordinate normal to the interface.

\begin{figure}[H]
	\centering
	\begin{subfigure}[t]{0.42\textwidth}
		\label{fig1a} 
		\includegraphics[width=\textwidth]{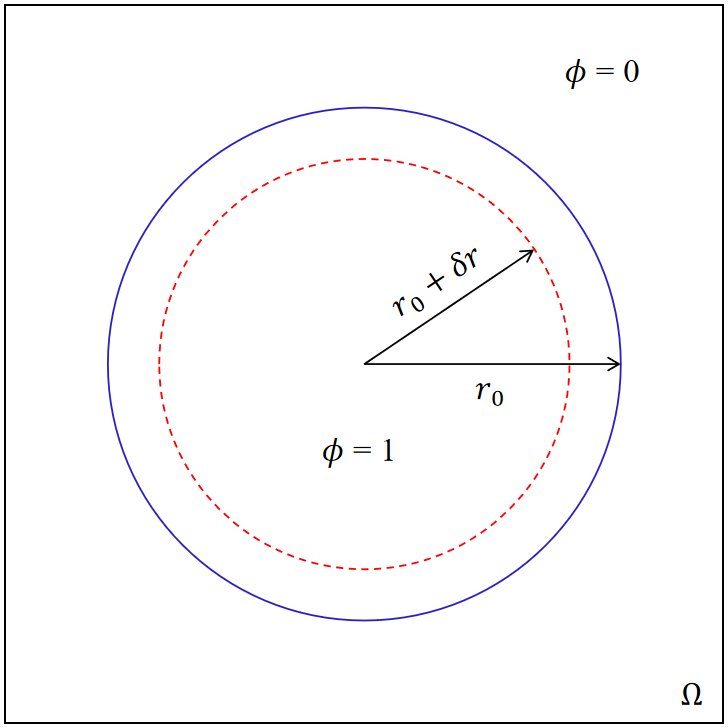}
		\caption{}
	\end{subfigure}		
	\hfill
	\begin{subfigure}[t]{0.44\textwidth}
		\label{fig1b} 
		\includegraphics[width=\textwidth]{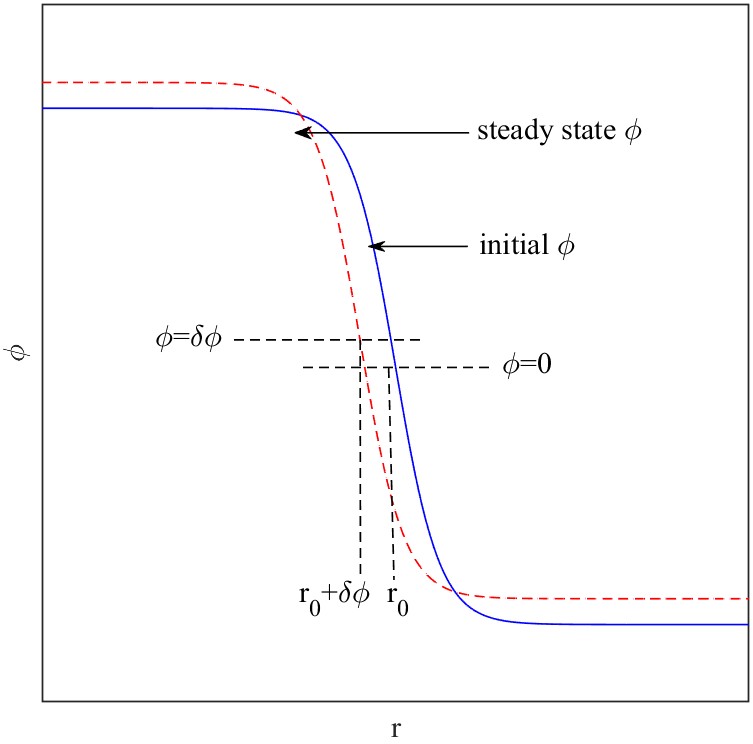}
		\caption{}
	\end{subfigure}
	\caption{Illustration of a circular droplet contracting within a stationary medium.} 
	\label{fig1}
\end{figure}

In recent years, the classical C-H equation (i.e. Eq. (\ref{eq3})) is widely used in different fields, such as spinodal decomposition \cite{Maraldi2012Aunified}, image inpainting \cite{Bertozzi2007Analysis}, the simulation of tumor growth \cite{Hilhorst2015Formal, Wise2008Three}, and the modeling of interfaces between distinct fluid phases \cite{Soligo2019Mass}, etc. However, it should be highlighted that Yue et al. \cite{Yue2007Spontaneous} indicated the solution to the C-H equation, while resembling the hyperbolic tangent profile, is not identical and is situated within the interval $[0+\xi, 1+\xi]$, and $\xi$ is a small value associated with interface thickness $W$. It is because the free energy is concentrated at the interface and the energy is zero in the bulk, while the bulk has a finite volume, the system may reduce the total energy by shrinking the region enclosed by the interface, and the order parameter is slightly deviate from its initial value (as depicted in Fig. \ref{fig1}).
Therefore, while the classical C-H model maintains the overall volume of the binary system, it fails to preserve the local volume. To address this issue, a straightforward approach is to ensure that the diffuse interface matches the hyperbolic tangent profile described in Eq. (\ref{eq4}). Based on this, Li et al. \cite{Li2016Aphase, Li2024local} inroduced an improved C-H equation, the core idea is to introduce an additional penalty energy $\varepsilon_p (\phi)$ term into the free energy. Specifically, when Eq. (\ref{eq4}) is satisfied, the normal derivative is articulated as
\begin{equation}
	|\nabla\phi|=\frac{d\phi}{ds}=\frac{4\phi(1-\phi)}{W}.
	\label{eq5}
\end{equation}
To ensure Eq. (\ref{eq4}) satisfied when minimizing the penalty energy, $\varepsilon_p (\phi)$ can be formulated as \cite{Li2016Aphase}
\begin{equation}
	\varepsilon_{p}(\phi)=\frac{\lambda}{2}\int_{\Omega}[\frac{4\phi(1-\phi)}{W}-|\nabla\phi|]^{2}d\Omega,
	\label{eq6}
\end{equation}
where $\lambda>0$ denotes the artificial penalty constant, which is set to 0.5 here \cite{Li2024local}. Then, the improved C-H equation is derived from the total energy given by
\begin{equation}
	\begin{aligned}
		\mathcal{F}_{total} & =\mathcal{F}_{\phi}+\varepsilon_{p}(\phi) \\
		& =\int_{\Omega}(\psi_{0}+\frac{\kappa}{2}|\nabla\phi|^{2})d\Omega+\frac{\lambda}{2}\int_{\Omega}[\frac{4\phi(1-\phi)}{W}-|\nabla\phi|]^{2}d\Omega.
	\end{aligned}
    \label{eq7}
\end{equation}
The gradient flows of free energy $\mathcal{F}_{\phi}$ and penalty energy $\varepsilon_{p}(\phi)$ under the $H^{-1}$ and $L_2$ norms are applied \cite{Bertozzi2006Inpainting}, respectively, and the results are substituted into the original C-H equation, yielding \cite{Li2016Aphase}
\begin{equation}
	\frac{\partial\phi}{\partial t}+\nabla\cdot(\phi\mathbf{u})=M\nabla^2\frac{\delta\mathcal{F}_\phi}{\delta\phi}-M\frac{\delta\varepsilon_p(\phi)}{\delta\phi}.
	\label{eq8}
\end{equation}
In the equation presented above, the initial term on the right side ${{\delta \mathop {\cal F}\nolimits_\phi  } \mathord{\left/ {\vphantom {{\delta \mathop {\cal F}\nolimits_\phi  } {\delta \phi }}} \right.\kern-\nulldelimiterspace} {\delta \phi }}$ corresponds to Eq. (\ref{eq2}), and another term ${{\delta \mathop \varepsilon \nolimits_p (\phi )} \mathord{\left/ {\vphantom {{\delta \mathop \varepsilon \nolimits_p (\phi )} {\delta \phi }}} \right.\kern-\nulldelimiterspace} {\delta \phi }}$ is obtained through the variational derivation of Eq. (\ref{eq6})
\begin{equation}
	\frac{\delta\varepsilon_{p}(\phi)}{\delta\phi}=\frac{\lambda(1-2\phi)}{4W}(\frac{4\phi(1-\phi)}{W}-|\nabla\phi|)-\lambda\nabla\cdot[\nabla\phi-\frac{4\phi(1-\phi)}{W}\frac{\nabla\phi}{|\nabla\phi|}].
	\label{eq9}
\end{equation}
In particular, for the first term in the above equation, as the interface profile converges to a hyperbolic tangent function, $4\phi(1-\phi)/W-|\nabla\phi|\to0$, then Eq. (\ref{eq9}) is approximated as
\begin{equation}
	\begin{split}
		\frac{\delta\varepsilon_{p}(\phi)}{\delta\phi} &\approx-\lambda\nabla\cdot[\nabla\phi-\frac{4\phi(1-\phi)}{W}\frac{\nabla\phi}{|\nabla\phi|}]\\
		&\approx-\lambda[\nabla^{2}\phi-\nabla\cdot(\frac{4\phi(1-\phi)}{W}\frac{\nabla\phi}{|\nabla\phi|})].
	\end{split}
	\label{eq10}
\end{equation}
By inserting Eqs. (\ref{eq2}) and (\ref{eq10}) into Eq. (\ref{eq8}), an improved C-H equation is derived as
\begin{equation}
	\frac{\partial\phi}{\partial t}+\nabla\cdot(\phi \mathbf{u}) = \nabla\cdot(M\nabla\hbar)+\lambda M[\nabla^{2}\phi-\nabla\cdot(\frac{4\phi(1-\phi)}{W}\frac{\nabla\phi}{|\nabla\phi|})].
	\label{eq11}
\end{equation}

Notice that, for the above C-H equation, the nonlinear sharpening flux $\phi (1-\phi)$ introduces jumps or discontinuities at the interface, especially when the interface thickness is thin, which can significantly impact precision loss in numerical simulations \cite{Liang2023Lattice}. To overcome this limitation, Tomasz \cite{Tomasz2015Aconsistent} and Jain \cite{Jain2022Accurate} recently proposed an improved method, which replaces the order parameter $\phi$ with the signed distance function $\psi$ from the level set method, thereby deriving an improved interface equation. Specifically, in this method, $\psi$ is defined as $\psi=s(\phi)-s(\phi=0.5)$, and combined with Eq. (\ref{eq4}), the derivative of the order parameter with respect to $\psi$ is defined as:
\begin{equation}
	\frac{d\phi}{d\psi}=\frac{d\phi}{d s}=\frac{4\phi(1-\phi)}{W}.
	\label{eq12}
\end{equation}
Then, integrating the aforementioned equation yields the relationship between $\phi$ and $\psi$:
\begin{equation}
	\phi  = \frac{{\mathop e\nolimits^{\displaystyle (4\psi /W)} }}{{\mathop e\nolimits^{\displaystyle (4\psi /W)}  + 1}}=0.5[1+\rm{tanh}(\frac{2\psi}{W})],
	\label{eq13}
\end{equation}
\begin{equation}
	\psi=\frac{W}{4}ln(\frac{\phi}{1-\phi}).
	\label{eq14}
\end{equation}
Now, using the relation in Eqs. (\ref{eq13}) and (\ref{eq14}), we can get
\begin{equation}
	\phi(1-\phi)=0.25[1-\rm{tanh}^{2}(\frac{2\psi}{W})].
	\label{eq15}
\end{equation} 
Additionally, we can further prove that:
\begin{equation}
	\frac{{\nabla \phi }}{{|\nabla \phi |}} = \frac{{\nabla \psi }}{{|\nabla \psi |}}.
	\label{eq16}
\end{equation}
\begin{figure}[H]
	\centering
	\includegraphics[width=0.55\textwidth]{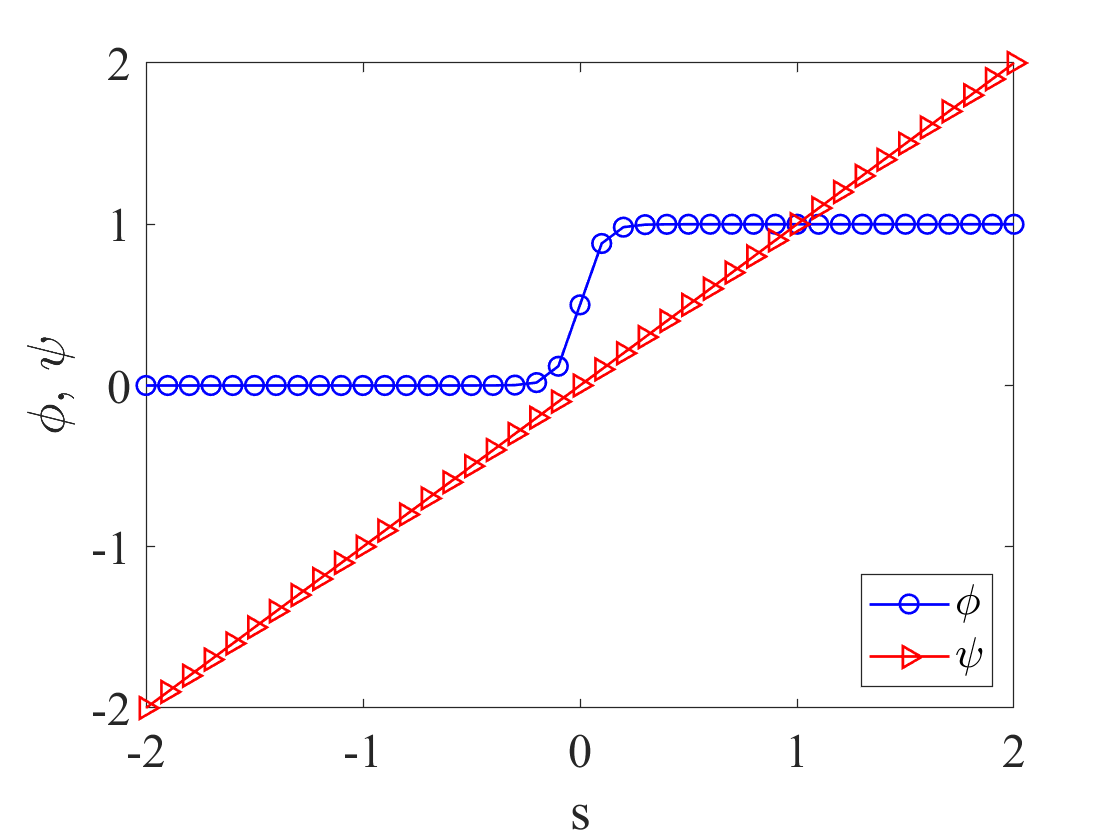}				
	\caption{A depiction of the variable $\phi$ and the signed distance function $\psi$}
	\label{fig2}
\end{figure}
Following previous works \cite{Tomasz2015Aconsistent, Chiodi2017Areformulation, Chiu2019Acoupled, Jain2022Accurate}, we modify the interface-capturing model by utilizing the relationship defined in Eq. (\ref{eq15}) to substitute the non-linear sharpening flux term $\phi (1-\phi)$ from Eq. (\ref{eq11}) with $\psi$. As shown in Fig. \ref{fig2}, the linearly varying function $\psi$ is inherently continuous and smooth across the interface, and is better behaved than $\phi$, which often has sharp gradients or discontinuities. Moreover, for the same reason, we can substitute ${{\nabla \phi } \mathord{\left/ {\vphantom {{\nabla \phi } {|\nabla \phi |}}} \right.\kern-\nulldelimiterspace} {|\nabla \phi |}}$ in Eq. (\ref{eq11}) using a formulation based on $\psi$ given by Eq. (\ref{eq16}). As a result, the improved C-H equation is ultimately derived \cite{Li2024local}:
\begin{equation}
	\frac{{\partial \phi }}{{\partial t}} + \nabla  \cdot (\phi {\bf{u}}) = \nabla  \cdot (M\nabla \hbar ) + \lambda M[\mathop \nabla \nolimits^2 \phi  - \nabla  \cdot \{ \frac{1}{W}[1 - \rm{tanh}^2(\frac{{2\psi }}{W})] \times \frac{{\nabla \psi }}{{|\nabla \psi |}}\} ].
	\label{eq17}
\end{equation}

For two-phase diffusive interfacial fluid modeling, the hydrodynamic equations also need to be considered. Assuming that the two fluids are incompressible and immiscible , with the gravitational effect being disregarded, the fluid flow can be controlled by the hydrodynamic equations \cite{Unverdi1992Afront}
\begin{equation}
	\begin{aligned}
		\nabla\cdot \mathbf{u}& =0, \\
		\partial_t(\rho\mathbf{u})+\nabla\cdot(\rho\mathbf{uu})& =-\nabla p+\nabla\cdot\left[\mu(\nabla \mathbf{u}+\nabla \mathbf{u}^{T})\right] +\mathbf{F}_s,
	\end{aligned}
    \label{eq18}
\end{equation}
in which $\rho$, $\bm{u}$, $p$ and $\mu$ represent the mass density, velocity, hydrodynamic pressure and fluid dynamics viscosity, respectively. The term $\mathbf{F}_s$ is the surface tension force, defined as $\mathbf{F}_s=\hbar\nabla\phi$, in which $\hbar$ indicates the chemical potential.

\section{Lattice Boltzmann method}
In this section, the LB method for simulating two-phase diffusive interfacial fluids is presented. In general, the LB methods are classified into three categories depending on the collision operator, such as the single-relaxation-time (SRT) model (also known as the Bhatnagar-Gross-Krook model, BGK) \cite{Qian1992Lattice}, the two-relaxation-time (TRT) model \cite{Ginzburg2008Two}, and the multiple-relaxation-time (MRT) model \cite{Lallemand2000Theory}. In this work, we will focus on the MRT model due to its superiority in stability and accuracy. The MRT model is performed in a regular Cartesian mesh with D$n$Q$q$ ($n$-dimensional $q$-velocity) lattice. For the two-dimensional scenarios addressed here, the D2Q9 discrete lattice velocities are utilized, as described in \cite{Krüger2017The}
\begin{equation}
	\left.\mathbf{c}_{i}=\left\{
	\begin{array}
		{ll}c\left(0,0\right), & i=0 \\
		c\left(\cos\left[\left(i-1\right)\pi/2\right],\sin\left[\left(i-1\right)\pi/2\right]\right), & i=1,2,3,4 \\
		\sqrt{2}c\left(\cos\left[\left(2i-9\right)\pi/4\right],\sin\left[\left(2i-9\right)\pi/4\right]\right), & i=5,6,7,8.
	\end{array}\right.\right.
    \label{eq19}
\end{equation}
Here, $i$ denotes the direction of the lattice, while $c = \Delta x/\Delta t$ signifies the speed of the lattice, where $\Delta x$ and $\Delta t$ represent the latice spacing and time step, respectively. Furthermore, within the D2Q9 framework, the sound speed $c_s$ is given by $c_s=c/{\sqrt 3}$, while the weight coefficients for various lattice directions are specified as $\omega_0 = 4/9, \omega_{1 - 4} = 1/9, \omega_{5 - 8} = 1/36$.

\subsection{The MRT Lattice Boltzmann equation for hydrodynamic equations}
The generalized evolution equation for incompressible fluid flow is expressed as \cite{Liang2014Phase}
\begin{equation}
	f_i(\mathbf{x}+\mathbf{c}_i\Delta t,t+\Delta t)=f_i(\mathbf{x},t)-(\mathbf{M}^{-1}\mathbf{S}^h\mathbf{M})_{ij}[f_j(\mathbf{x},t)-f_j^{eq}(\mathbf{x},t)]+\Delta t(1-\frac{1}{2\tau_f})F_i(\mathbf{x},t),
	\label{eq20}
\end{equation}
in which $f_i(\mathbf{x},t)$ represents the distribution function at position $\mathbf{x}$ and time $t$, and $F_i(\mathbf{x},t)$ denotes the discrete forcing term given by
\begin{equation}
	F_i(\boldsymbol{x},t)=\omega_i[\mathbf{u}\cdot\nabla\rho+\frac{\mathbf{c}_i\cdot\mathbf{F}}{c_s^2}+\frac{\mathbf{u}\nabla\rho:(\mathbf{c}_i\mathbf{c}_i-c_s^2\mathbf{I})}{c_s^2}],
	\label{eq21}
\end{equation}
in which $\mathbf{I}$ stands for the unit matrix. The local equilibrium distribution function $f_i^{eq} (\mathbf{x},t)$ in Eq. (\ref{eq20}) can be expressed as
\begin{equation}
	f_i^{e q}(\mathbf{x}, t)=\left\{\begin{array}{cc}
		\left(\omega_0-1\right) \frac{p}{c_s^2}+\rho s_i(\mathbf{u}), & i=0, \\
		\omega_i \frac{p}{c_s^2}+\rho s_i(\mathbf{u}), & i \neq 0,
	\end{array}\right.
	\label{eq22}
\end{equation}
with
\begin{equation}
	s_i(\mathbf{u})=\omega_i\left[\frac{\mathbf{c}_i \cdot \mathbf{u}}{c_s^2}+\frac{\left(\mathbf{c}_i \cdot \mathbf{u}\right)^2}{2 c_s^4}-\frac{\mathbf{u} \cdot \mathbf{u}}{2 c_s^2}\right].
	\label{eq23}
\end{equation}
In addition, the matrix $\mathbf{M}$ in Eq. (\ref{eq20}) serves as the transformation matrix, as specified in \cite{Chai2016Amultiple}
\begin{equation}
	\mathbf{M}=\left[\begin{array}{ccccccccc}
		1 & 1 & 1 & 1 & 1 & 1 & 1 & 1 & 1 \\
		-4 & -1 & -1 & -1 & -1 & 2 & 2 & 2 & 2 \\
		4 & -2 & -2 & -2 & -2 & 1 & 1 & 1 & 1 \\
		0 & 1 & 0 & -1 & 0 & 1 & -1 & -1 & 1 \\
		0 & -2 & 0 & 2 & 0 & 1 & -1 & -1 & 1 \\
		0 & 0 & 1 & 0 & -1 & 1 & 1 & -1 & -1 \\
		0 & 0 & -2 & 0 & 2 & 1 & 1 & -1 & -1 \\
		0 & 1 & -1 & 1 & -1 & 0 & 0 & 0 & 0 \\
		0 & 0 & 0 & 0 & 0 & 1 & -1 & 1 & -1
	\end{array}\right].
    \label{eq24}
\end{equation}
Notice that multiplying Eq. (\ref{eq20}) by the transformation matrix $\mathbf{M}$ projects $f_i (\mathbf{x},t)$, $f_i^{eq} (\mathbf{x},t)$ and $F_i (\mathbf{x},t)$ into the moment space, resulting in $\mathbf{m}_f (\mathbf{x},t)=\mathbf{M} f (\mathbf{x},t)$, $\mathbf{m}_f^{eq} (\mathbf{x},t)=\mathbf{M} f^{eq} (\mathbf{x},t)$ and $\overline{\mathbf{F}}(\mathbf{x},t)=\mathbf{M} F(\mathbf{x},t)$, which effectively shifts the collision process from velocity space to momentum space, as shown in the equation:
\begin{equation}
	\mathbf{m}_f^*(\mathbf{x}, t)=\mathbf{m}_f(\mathbf{x}, t)-\mathbf{S}^f\left[\mathbf{m}_f(\mathbf{x}, t)-\mathbf{m}_f^{e q}(\mathbf{x}, t)\right]+\Delta t\left(\mathbf{I}-\frac{\mathbf{S}^f}{2}\right) \overline{\mathbf{F}}(\mathbf{x}, t),
	\label{eq25}
\end{equation}
in which $\mathbf{S}^f$ represents a diagonal relaxation matrix defined as 
\begin{equation}
	\mathbf{S}^f=\operatorname{diag}\left(s_0^f, s_1^f, s_2^f, s_3^f, s_4^f, s_5^f, s_6^f, s_7^f, s_8^f\right),
	\label{eq26}
\end{equation}
with the constraint $0<s_i^f<2$. If all $s_i^f$ are identical, the MRT model simplifies to the SRT model. Additionally, the relaxation time $\tau_f$ correlates with the dynamic viscosity $\mu$ and density $\rho$ by the following relationship:
\begin{equation}
	\mu=\rho c_s^2\left(\tau_f-\frac{1}{2}\right)\Delta t,
	\label{eq27}
\end{equation}
where $\tau_f=1/s_7^f=1/s_8^f$. Using the distribution function, one can calculate the velocity $\mathbf{u}$ and the hydrodynamic pressure $p$ as
\begin{equation}
	\rho \mathbf{u}=\sum_i \mathbf{c}_i f_i+\frac{1}{2} \Delta t \mathbf{F},
	\label{eq28}
\end{equation}
\begin{equation}
	p=\frac{c_s^2}{1-\omega_0}\left[\sum_{i \neq 0} f_i+\frac{\Delta t}{2} \mathbf{u} \cdot \nabla \rho+\rho s_0(\mathbf{u})\right] .
	\label{eq29}
\end{equation}

\subsection{The MRT Lattice Boltzmann equation for modified Cahn-Hilliard equation}
Inspired by previous work of LB model for incompressible multiphase flows \cite{Liang2023Lattice, Liang2014Phase}, we have derived a corresponding evolution equation to match the target C-H equation (i.e., Eq. (\ref{eq17})). The evolution equation can be formulated as
\begin{equation}
	g_i\left(\mathbf{x}+\boldsymbol{c}_i \Delta t, t+\Delta t\right)=g_i(\mathbf{x}, t)-\left(\mathbf{M}^{-1} \mathbf{S}^g \mathbf{M}\right)_{i j}\left[g_j(\mathbf{x}, t)-g_j^{e q}(\mathbf{x}, t)\right]+\Delta t\left(1-\frac{1}{2 \tau_g}\right) G_i(\mathbf{x}, t),
	\label{eq30}
\end{equation}
where $g_i^{e q}(\mathbf{x}, t)$ is the local equilibrium distribution function,  specified as
\begin{equation}
	g_i^{e q}(\mathbf{x}, t)=\left\{\begin{array}{cc}
		\phi+\left(\omega_i-1\right)(\hbar+\lambda \phi) & i=0, \\
		\omega_i(\hbar+\lambda \phi)+\omega_i \frac{\mathbf{c}_i \cdot \phi \mathbf{u}}{c_s^2} & i \neq 0.
	\end{array}\right.
    \label{eq31}
\end{equation}
In Eq. (\ref{eq17}), the diffusion term with $\psi$ serves as the source term. Accounting for the discrete lattice effect, we develop an effective forced distribution function accordingly as
\begin{equation}
	G_i(\mathbf{x},t)=\omega_i\boldsymbol{c}_i\cdot\{\frac{\partial_t(\phi\mathbf{u})}{c_s^2}+\frac{\lambda}{W}[1-\rm{tanh}^2(\frac{2\psi}{W})]\times\frac{\nabla\psi}{|\nabla\psi|}\}.
	\label{eq32}
\end{equation}
Analogous to the transformation applied in Eq. (\ref{eq25}), the evolution Eq. (\ref{eq30}) can be converted into the momentum space through multiplication by the matrix $\mathbf{M}$, yielding:
\begin{equation}
	\mathbf{m}_g^*(\mathbf{x},t)=\mathbf{m}_g(\mathbf{x},t)-\mathbf{S}^g[\mathbf{m}_g(\mathbf{x},t)-\mathbf{m}_g^{eq}(\mathbf{x},t)]+\Delta t(\mathbf{I}-\frac{\mathbf{S}^g}{2})\mathbf{\overline{G}}(\mathbf{x},t),
	\label{eq33}
\end{equation}
in which $\mathbf{m}_g(\mathbf{x},t)=\mathbf{M} g(\mathbf{x},t)$, $\mathbf{m}_g^{eq}(\mathbf{x},t)=\mathbf{M} g^{eq}(\mathbf{x},t)$, $\mathbf{\overline{G}}(\mathbf{x},t)=\mathbf{M}G_i(\mathbf{x},t)$. Additionally, the mobility $M$ is represented as a function of the relaxation time $\tau_g$
\begin{equation}
	M = \mathop c\nolimits_s^2 (\mathop \tau \nolimits_g  - \frac{1}{2})\Delta t,
	\label{eq34}
\end{equation}
where $\tau_g=1/s_3^g=1/s_5^g$. The order parameter $\phi$ and the signed-distance function $\psi$ are calculated by
\begin{equation}
	\phi  = \sum\limits_i {\mathop g\nolimits_i }, 
	\label{eq35}
\end{equation}
\begin{equation}
	\psi  = \frac{W}{4}\ln (\frac{\phi }{{1 - \phi }}).
	\label{eq36}
\end{equation}

It must be emphasized that the accurate calculation of $\psi$ is crucial for the realization of the LB model of the modified C-H equation. However, according to Eq. (\ref{eq36}), the value of $\psi$ tends to infinity in the bulk phase region (i.e., $\phi=0$ or 1). To address this singularity issue in advance, it is advisable to introduce a minimal value in the computation,
\begin{equation}
	\psi  = \frac{W}{4}\ln (\frac{\phi + \xi}{{1 - \phi +\xi }}).
	\label{eq37}
\end{equation}
Specifically, setting $\xi=10^{-10}$ in simulations to ensure numerical stability and continuity. Additionally, due to the slight diffusion effects, the order parameter $\phi$ may occasionally fall outside the range $[0,1]$, resulting in unphysical values of $\psi$ from Eq. (\ref{eq37}). To resolve this, $\phi$ is clipped within $[0,1]$ before calculating $\psi$, that is using $\phi_{clip}=min[max(0, \phi), 1]$ as the adjusted value instead of $\phi$, which does not affect data preservation since $\phi$ is not directly updated but only modified temporarily for the computation of $\psi$ \cite{Liang2023Lattice}.

For efficient numerical computation, the evolution equations' derivative terms need to be discretized through suitable numerical techniques. In this study, a backward difference scheme is employed to calcuate the time derivatives \cite{Liang2023Lattice}
\begin{equation}
	\mathop \partial \nolimits_t (\phi \mathbf{u}) = \frac{{\phi (t)\mathbf{u}(t) - \phi (t - \Delta t)\mathbf{u}(t - \Delta t)}}{{\Delta t}}.
\end{equation}
Finally, in order to preserve the numerical precision of the LB model, a classical second-order isotropic discretization scheme is implemented for computing the gradient and Laplace operator terms \cite{Thampi2013Isotropic}
\begin{equation}
	\nabla \boldsymbol{\chi}  = \sum\limits_{i \ne 0} {\frac{{\mathop \omega \nolimits_i \mathop \mathbf{c}\nolimits_i \boldsymbol{\chi} (\mathbf{x} + \mathop \mathbf{c}\nolimits_i \Delta t)}}{{\mathop c\nolimits_s^2 \Delta t}}},
\end{equation}
\begin{equation}
	\mathop \nabla \nolimits^2 \boldsymbol{\chi}  = \sum\limits_{i \ne 0} {\frac{{\mathop {2\omega }\nolimits_i [\boldsymbol{\chi} (\mathbf{x} + \mathop \mathbf{c}\nolimits_i \Delta t) - \boldsymbol{\chi} (\mathbf{x})]}}{{\mathop c\nolimits_s^2 \mathop {\Delta t}\nolimits^2 }}} ,
\end{equation}
where $\boldsymbol{\chi}$ being any physical variable.

\section{Results and discussion}
In this section, we evaluate the precision and reliability of the current LB model for the modified C-H equation by simulating two stationary droplets immersed in the gas phase \cite{Li2024local}, single vortex \cite{Bao2024Phase, Liang2023Lattice}, Rayleigh-Plateau fluid instability \cite{Li2024local, Li2016Aphase}, and droplet deformation under a shear flow problems \cite{Lee2020Effect}, and compared with the results from the LB model in Ref. \cite{Liang2014Phase}. We are primarily concerned with its performance in maintaining local volume conservation as well as interface preservation.

\subsection{Two stationary droplets immersed in the gas phase}
We first illustrate the current model's ability to maintain local volume conservation by simulating two stationary droplets suspended in the gas phase \cite{Li2024local}. We consider two fixed and separated stationary droplets located at the base of a computational domain measuring $2L \times L = 200 \times 100$. The initial setup for the profile of the droplets is given by
\begin{equation}
	\phi  = 0.5[1 + \tanh (\frac{{2(\mathop r\nolimits_1  - \sqrt {\mathop {(x - 80)}\nolimits^2  + \mathop y\nolimits^2 } )}}{W})] + 0.5[1 + \tanh (\frac{{2(\mathop r\nolimits_2  - \sqrt {\mathop {(x - 160)}\nolimits^2  + \mathop y\nolimits^2 } )}}{W})] .
\end{equation}
In this simulation, the original velocity and pressure are equal to 0, with an interface thickness of $W=3.0$. In addition, the following parameters are adopted: density ratio ($\rho_l/\rho_g)$ is 1.0, the dynamics viscosity ratio ($\mu_l / \mu_g)$ is 1.0, the mobility ($M$) is 0.1, and the surface tansion ($\sigma$) is 0.01. (surface tansion), unless otherwise stated.
\begin{figure}[H]
	\centering
	\begin{subfigure}[t]{0.56\textwidth}
		\includegraphics[width=\textwidth]{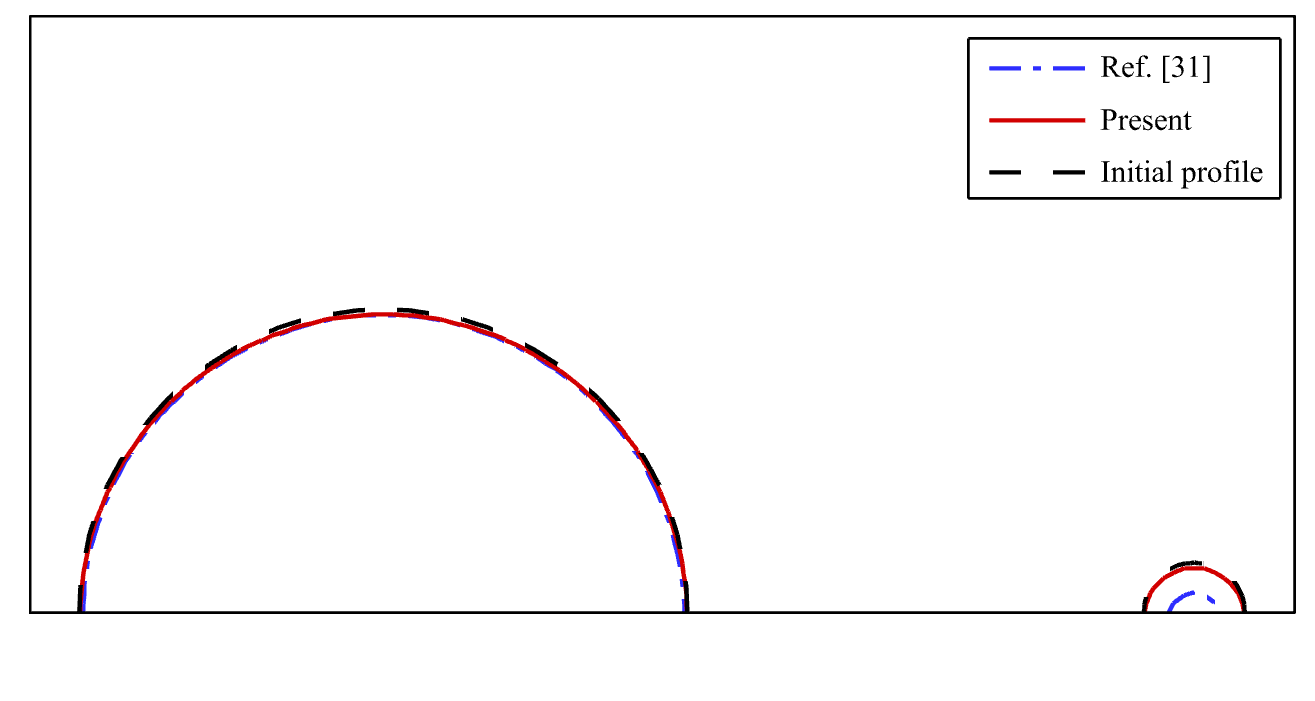}
		\caption{}
		\label{fig3a} 
	\end{subfigure}		
	\hfill
	\begin{subfigure}[t]{0.42\textwidth}
		\includegraphics[width=\textwidth]{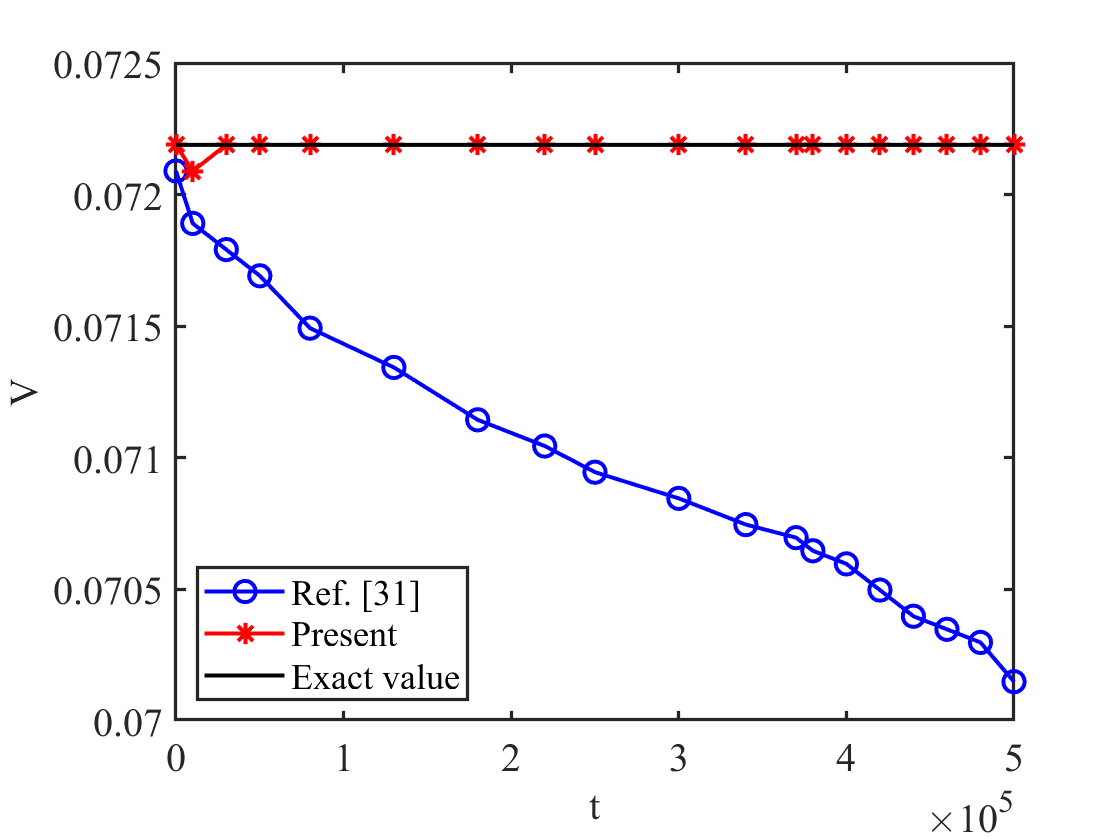}
		\caption{}
		\label{fig3b} 
	\end{subfigure}
	\caption{Comparison of the current model with the classical C-H model for local volume preservation, (a) interface profile configuration, (b) droplets local volume evolution over time.} 
	\label{fig3}
\end{figure}

Fig. (\ref{fig3a}) displays the interface profiles of the current model and the LB model in Ref. \cite{Liang2014Phase} at a specific in time. From the observations, the droplets calculated by the LB model for classical C-H equation in Ref. \cite{Liang2014Phase} exhibit obvious contraction, while the interface profile obtained by the current model shows better performance, particularly the difference in the simulation of small droplets is more obvious. Specifically, to assess the local volume discrepancy between the two models in a quantitative manner, Fig. (\ref{fig3b}) illustrates the evolution of droplets volume of the current model as well as the LB model for the classical C-H equation in Ref. \cite{Liang2014Phase} over time, where the black line depicted in the figure indicates the initial droplet volume. The results show that the volume of droplets calculated by the current model remains nearly constant. In contrary, the local volume calculated using the classical C-H model decreases significantly over time.

\subsection{Single vortex}
To further assess the current model's ability to address complex challenges with deforming interfaces, we will conduct a canonical numerical simulation of a time-reversed single vortex \cite{Liang2023Lattice}, and it is widely used as one of the benchmark test problems for accessing the model's capability in capturing the interface between two phases. The fundamental principle involves a given complex velocity field causing a circular interface to deform and elongate. Afterward, the interface goes back to its original configuration over a certain period of time, with significant changes in the interface topology during this process. In this simulation, a disk is positioned centrally above a square computational domain with a grid resolution of $L \times L=200 \times 200$. The disk has a radius of $R=45$, with its center located at $(0.5L,0.75L)$. The original shape of the disk is determined using the hyperbolic tangent function, which is specified as
\begin{equation}
	\phi {\rm{ = 0}}{\rm{.5[1 + tanh(}}\frac{{2(R - \sqrt {\mathop {(x - 0.5L)}\nolimits^2  + \mathop {(y - 0.75L)}\nolimits^2 } )}}{W}{\rm{)]}}.
\end{equation}
In addition, a complex nonlinear velocity field is taken in this simulation, the shear velocity with time is expressed as follows, \cite{Bao2024Phase},
\begin{equation}
	{\rm{u(x,y) = }}\mathop U\nolimits_0 \mathop {\sin }\nolimits^2 (\frac{{\pi x}}{L})\sin (\frac{{2\pi y}}{L})\cos (\frac{{\pi t}}{T}),
	\label{eq43}
\end{equation}
\begin{equation}
	{\rm{v(x,y) =  - }}\mathop U\nolimits_0 \mathop {\sin }\nolimits^2 (\frac{{\pi y}}{L})\sin (\frac{{2\pi x}}{L})\cos (\frac{{\pi t}}{T}),
	\label{eq44}
\end{equation}
where $T=nL/U_0$ periodic time with $n$ being a positive integer, and $t$ represents the evolutional time scaled by $L/U_0$. Notice that the topology of the interface has a larger deformation as we increase the periodic time $T$. However, most of the previous simulations using finite-difference, level-set method and LB method are primarily focused on short evolutionary stages with $n=2$ \cite{Tomasz2015Aconsistent, Geier2015Conservative} or $n=4$ \cite{Chiu2011Aconservative, Hao2024Aninterfacial}. Here, a longer periodic time $n=6$ is considered, with another physical parameter $U_0$ set to $0.04$.
\begin{figure}[H]
	\centering
	\begin{subfigure}[t]{0.76\textwidth}
		\includegraphics[width=\textwidth]{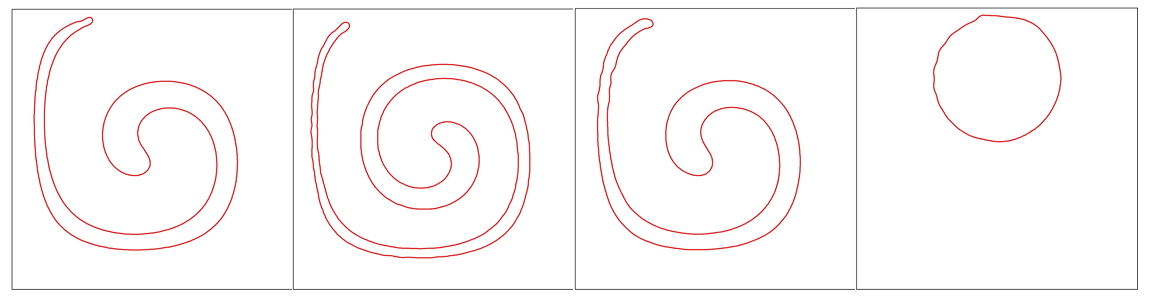}
		\caption{}
		\label{fig4a} 
	\end{subfigure}		
	\hfill
	\begin{subfigure}[t]{0.76\textwidth}
		\includegraphics[width=\textwidth]{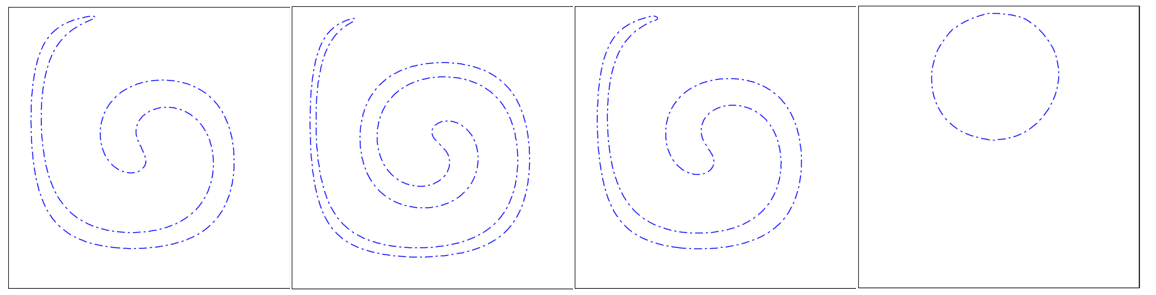}
		\caption{}
		\label{fig4b} 
	\end{subfigure}
	\caption{Interfacial shape profiles of the disk at different times, times in order $T/4$, $T/2$, $3T/4$, $T$, (a) the LB model in Ref. \cite{Liang2014Phase}, (b) the present model.} 
	\label{fig4}
\end{figure}

\begin{figure}[H]
	\centering
	\begin{subfigure}[t]{0.37\textwidth}
		\includegraphics[width=\textwidth]{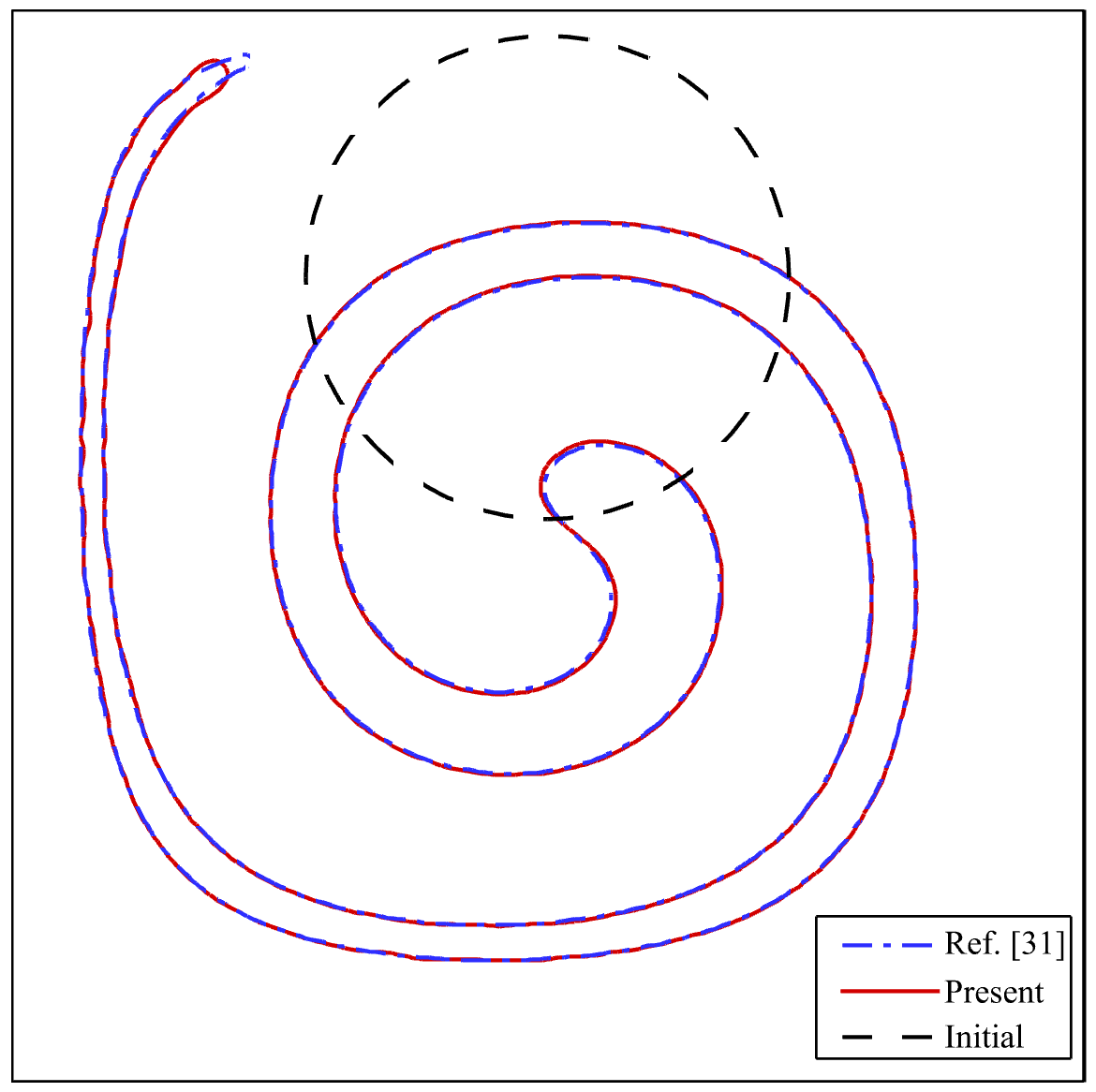}
		\caption{}
		\label{fig5a} 
	\end{subfigure}	
    \hspace{1cm}		
	\begin{subfigure}[t]{0.37\textwidth}
		\includegraphics[width=\textwidth]{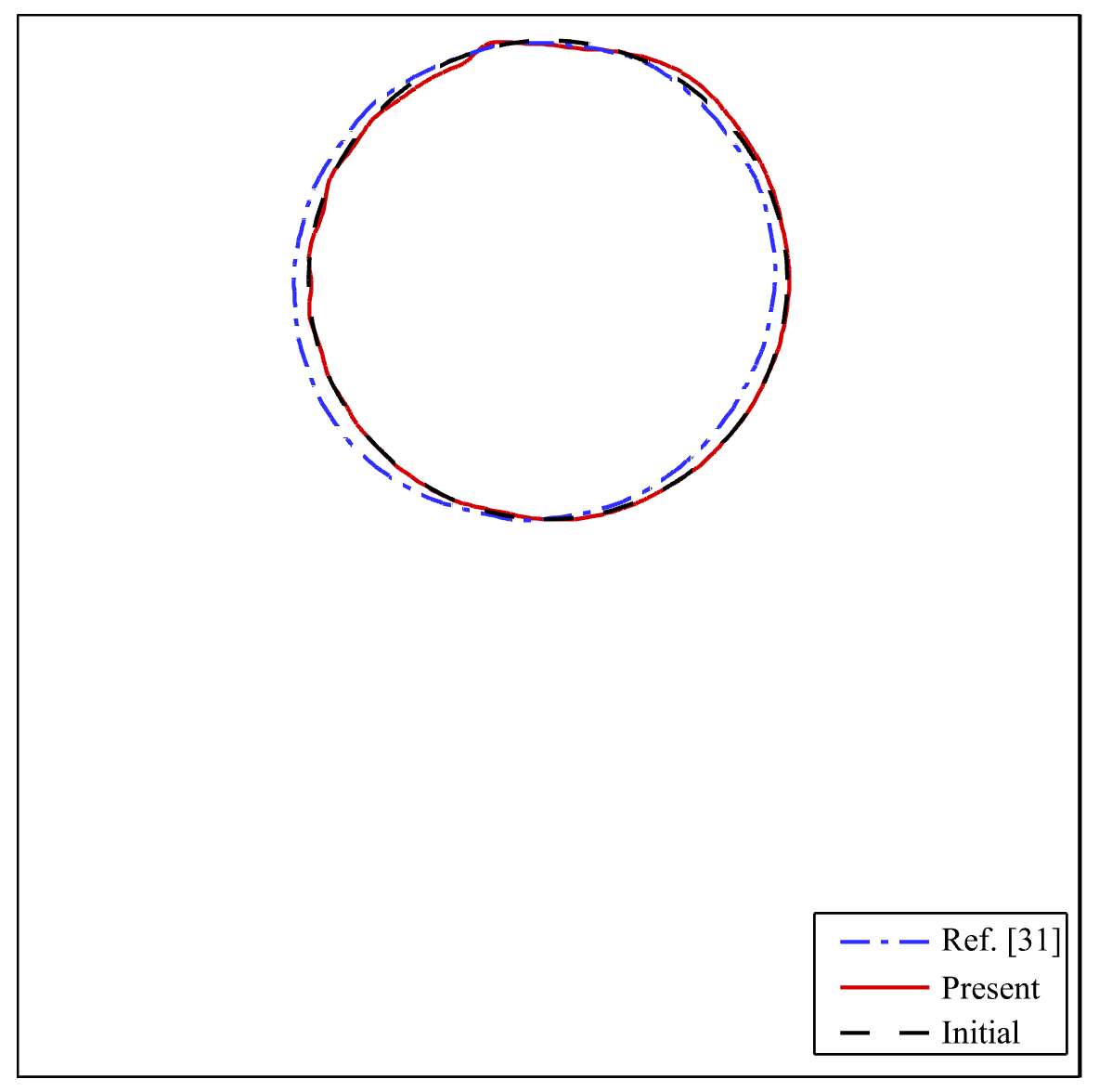}
		\caption{}
		\label{fig5b} 
	\end{subfigure}
	\caption{Comparison of the interface profiles between the present model and the LB model in Ref. \cite{Liang2014Phase} with the initial state at different times (a) $T/2$, (b) $T$.} 
	\label{fig5}
\end{figure}

Fig. \ref{fig4} demonstrates the shapes of the interfaces computed by the LB model in Ref. \cite{Liang2014Phase} as well as the current LB model at different stages within a single period, reaching the maximum degree of deformation occurs at $T/2$, where the direction of velocity reverses according to the time-reversal function specified by Eq. (\ref{eq43}) and (\ref{eq44}), ensuring the disk returns to the original configuration over completing a full period $T$. It is evident that the classical C-H model predicts significant diffusion near the interface at time $T$, while the interface captured by the current model performs better. Specifically, as shown in Fig. \ref{fig5}, slight differences in the interface shapes are noticeable between the classical and modified models at both $T/2$ and $T$. Particularly, at time T, the classical C-H model's prediction shows significant deformation and deviation compared to the initial state, while while the modified model maintains a more stable interface.

\subsection{Rayleigh-Plateau fluid instability}
Rayleigh-Plateau fluid instability is a type of interfacial instability caused by the interplay between surface tension and fluid inertia. In a two-phase fluid system consisting of liquid and gas, the surface tension drives the fluid interface to minimize its surface area under a given volume. When the interface of two immiscible fluids is subjected to external disturbances, the balance between surface tension and inertial forces is disrupted, resulting in irregular deformations of the interface. Such instability typically shows as a liquid column fragments into a series of satellite droplets to further reduce the surface area \cite{Li2024local}. The surface area of these small droplets is smaller than the original liquid column \cite{Li2016Aphase}, thereby reducing the total energy of the system. 

To demonstrate the advantages of present model in capturing small droplets effectively, we will conduct simulations of the Rayleigh-Plateau instability within a two-phase fluid framework. The original conditions is provided in \cite{Li2024local}
\begin{equation}
	\phi {\rm{ = 0}}{\rm{.5[1 + tanh(}}\frac{{2(a - y + 0.05\cos (x))}}{W}{\rm{)]}},
	\label{eq45}
\end{equation}
where the simulation is conducted within the designated domain $\Omega  = (0,2\pi ) \times (0,\pi )$ with mesh grids of $256 \times 128$. The parameter $a$ represents the initial thickness of the liquid column, and different initial thicknesses result in  varying outcomes in the evolution of the Rayleigh-Plateau fluid instability. In the simulation, $a$ is set to $0.15$, in addition, the original velocity and pressure are initialized to $0$. 

\begin{figure}[H]
	\centering
	\begin{subfigure}[t]{0.45\textwidth}
		\includegraphics[width=\textwidth]{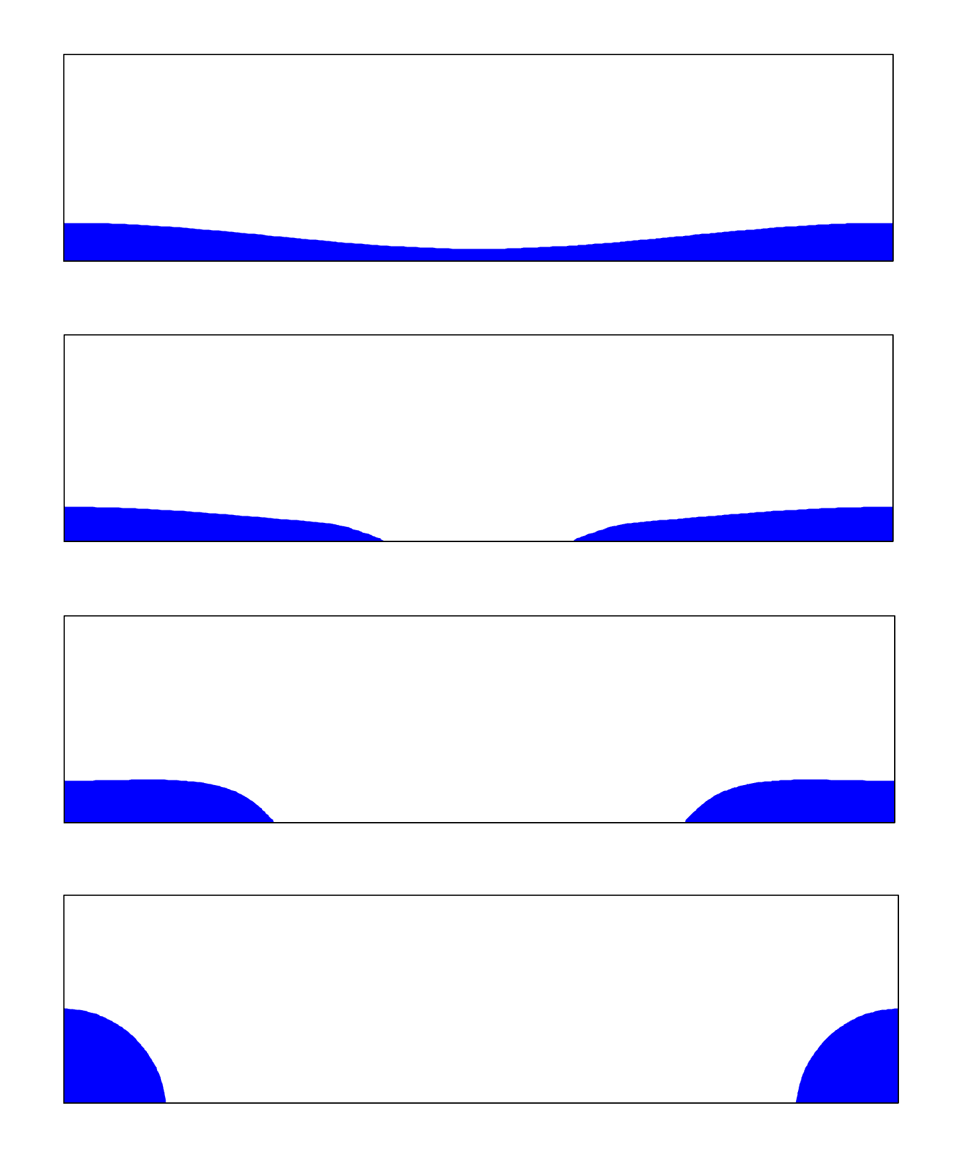}
		\caption{}
		\label{fig6a} 
	\end{subfigure}	
	\hspace{1cm}		
	\begin{subfigure}[t]{0.45\textwidth}
		\includegraphics[width=\textwidth]{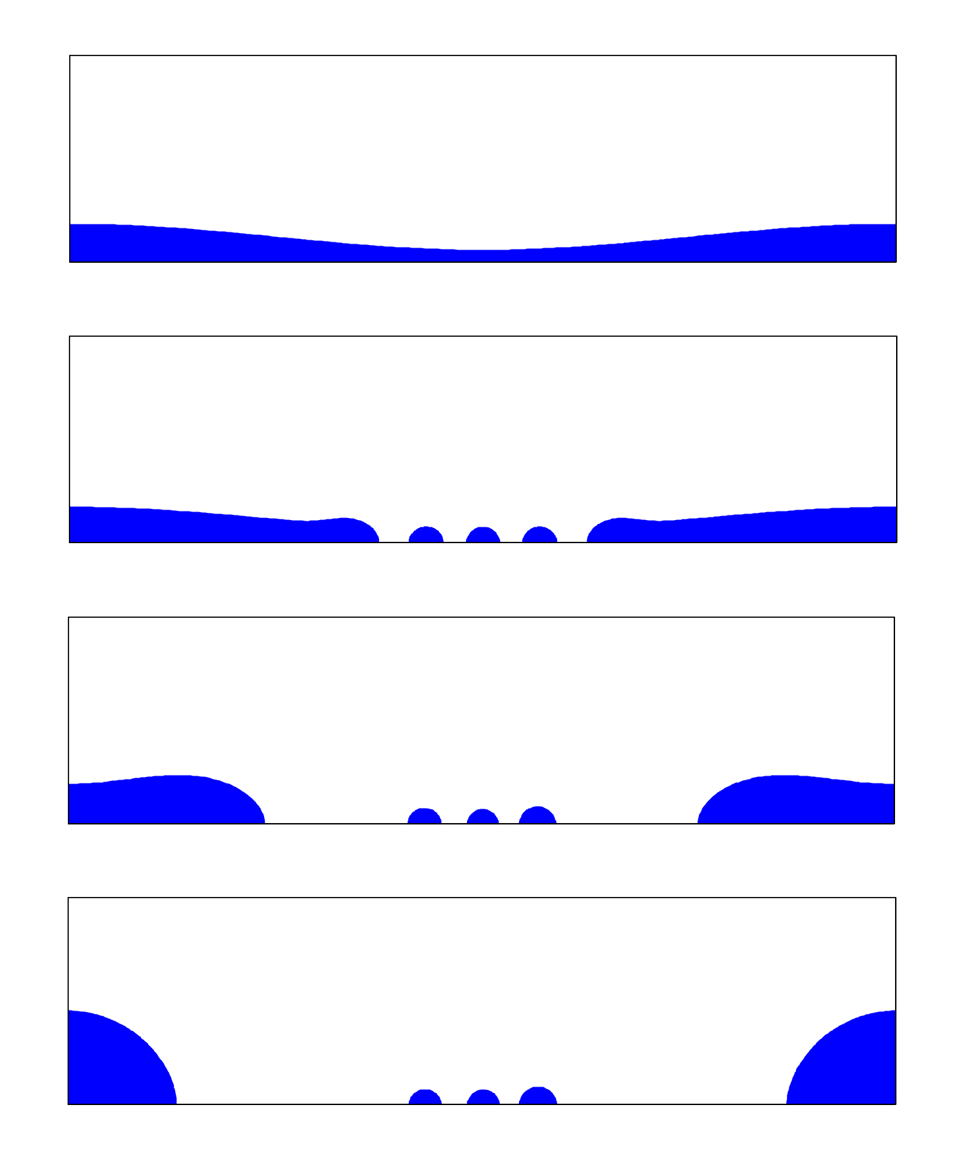}
		\caption{}
		\label{fig6b} 
	\end{subfigure}
	\caption{Profiles of Rayleigh-Plateau fluid instability, (a) LB model in Ref. \cite{Liang2014Phase}, (b) the present model.} 
	\label{fig6}
\end{figure} 

Figure 6 displays the interface profiles of the Rayleigh-Plateau fluid instability evolution for both the current model and the LB model from Ref. \cite{Liang2014Phase} at different times, respectively. According to the simulation results, with time evolution, the droplet symmetrically deviates from the median line of the computational domain ($x=\pi$) and shrinks toward both sides, giving rise to satellite droplets. The formation of satellite droplets arises from nonlinear terms in hydrodynamic equations. From the comparison between Fig. (\ref{fig6a}) and (\ref{fig6b}), it is apparent that the computed satellite droplets using the current model increase due to the surface tension and remain for a long time evolution, preserving more local volume and capturing the interface more accurately. In contrast, the satellite droplets computed by the LB model in Ref. \cite{Liang2014Phase} exhibit significant contraction and gradually disappear. It is crucial to highlight that the C-H equation primarily characterizes the phase separation process, where the two components of a binary fluid spontaneously separate and redistribute within their respective regions. Consequently, the classical C-H equation has limitations in maintaining the stability of small droplets and accurately capturing their dynamic behavior \cite{Li2016Aphase, Cahn1961Acta}.

\subsection{Droplet deformation under a shear flow}

\begin{figure}[H]
	\centering
	\includegraphics[width=0.55\textwidth]{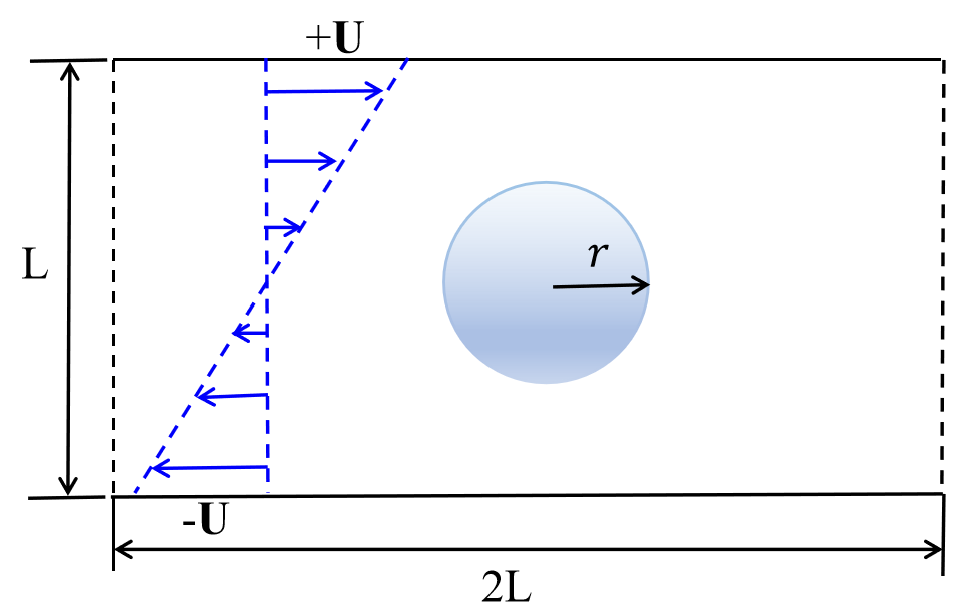}				
	\caption{Diagram of a droplet subjected to shear flow.}
	\label{fig7}
\end{figure}

Within the domain of phase field methods, volume diffusion often leads to contraction and coarsening phenomena. If the interface profile strays from its equilibrium condition, the volume within the interface will diffuse into the adjacent phase in an attempt to re-establish equilibrium, thereby triggering contraction. As noticed by Yue et al. \cite{Yue2007Spontaneous} and Lee et al. \cite{Lee2020Effect}, in the classical C-H model, there exists a critical droplet radius $r_c$. All droplets with a radius smaller than $r_c$ will eventually disappear through a process analogous to Ostwald ripening. To verify this phenomenon, we investigated a droplet evolution under the coarse grid and shear flow conditions, which is a canonical case for evaluating the precision and effectiveness of phase field methods in simulating two-phase fluid dynamics. The configuration is illustrated in Fig. \ref{fig7}, featuring a droplet of radius $r$ positioned centrally with a rectangular cavity. The cavity dimensions are $2L \times L$, with the upper and lower  parallel plates moving in opposite directions at a constant speed $U$, respectively, forming a shear flow. For this simulation, the computational domain spans $(0,128) \times (0,64)$, the droplet radius is $r=8$, and the initial droplet profile is as follow,
\begin{equation}
	\phi {\rm{ = 0}}{\rm{.5[1 + tanh(}}\frac{{2(R - \sqrt {\mathop {(x - 0.5L)}\nolimits^2  + \mathop {(y - 0.5L)}\nolimits^2 } )}}{W}{\rm{)]}}.
	\label{eq46}
\end{equation}

\begin{figure}[H]
	\centering
	\begin{subfigure}[t]{0.37\textwidth}
		\includegraphics[width=\textwidth]{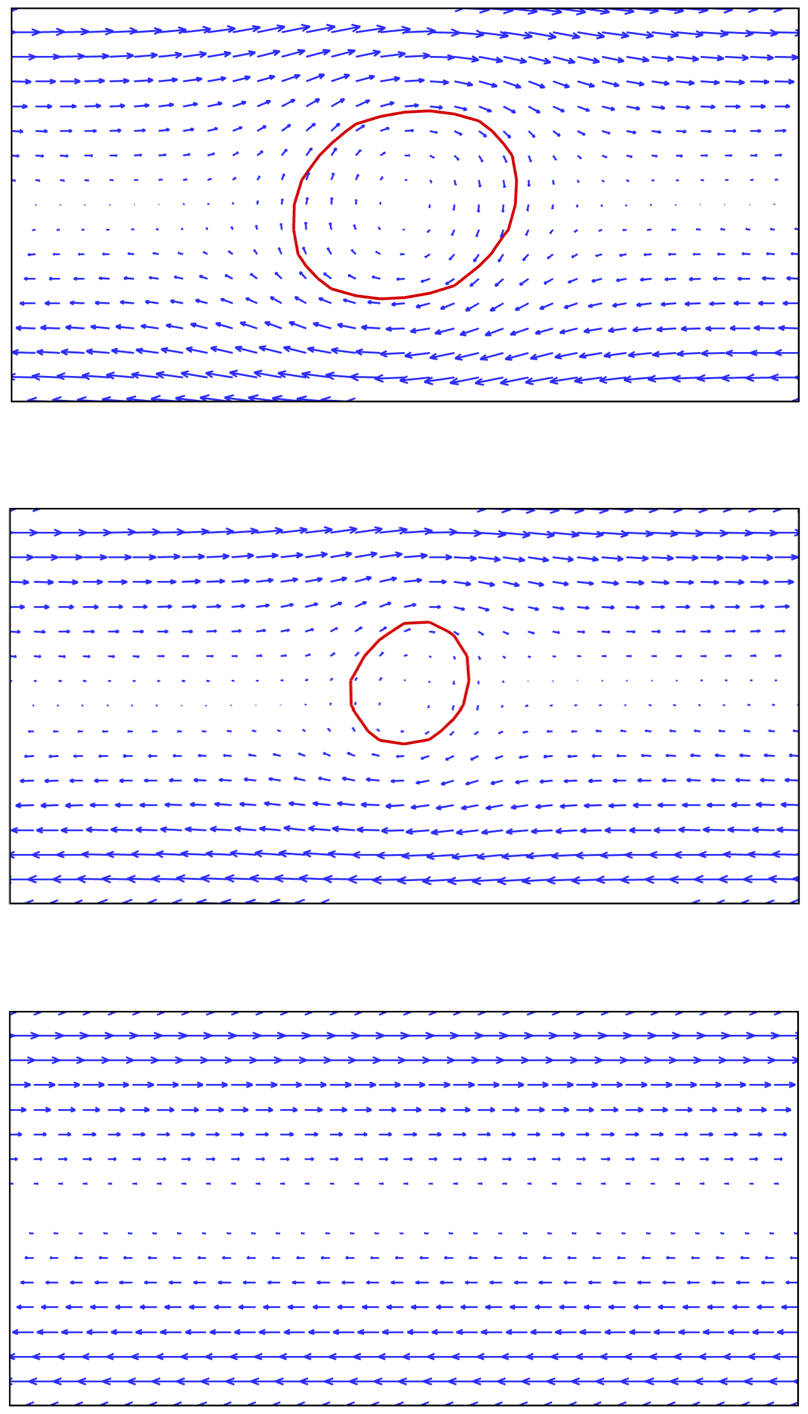}
		\caption{}
		\label{fig8a} 
	\end{subfigure}	
	\hspace{1cm}		
	\begin{subfigure}[t]{0.37\textwidth}
		\includegraphics[width=\textwidth]{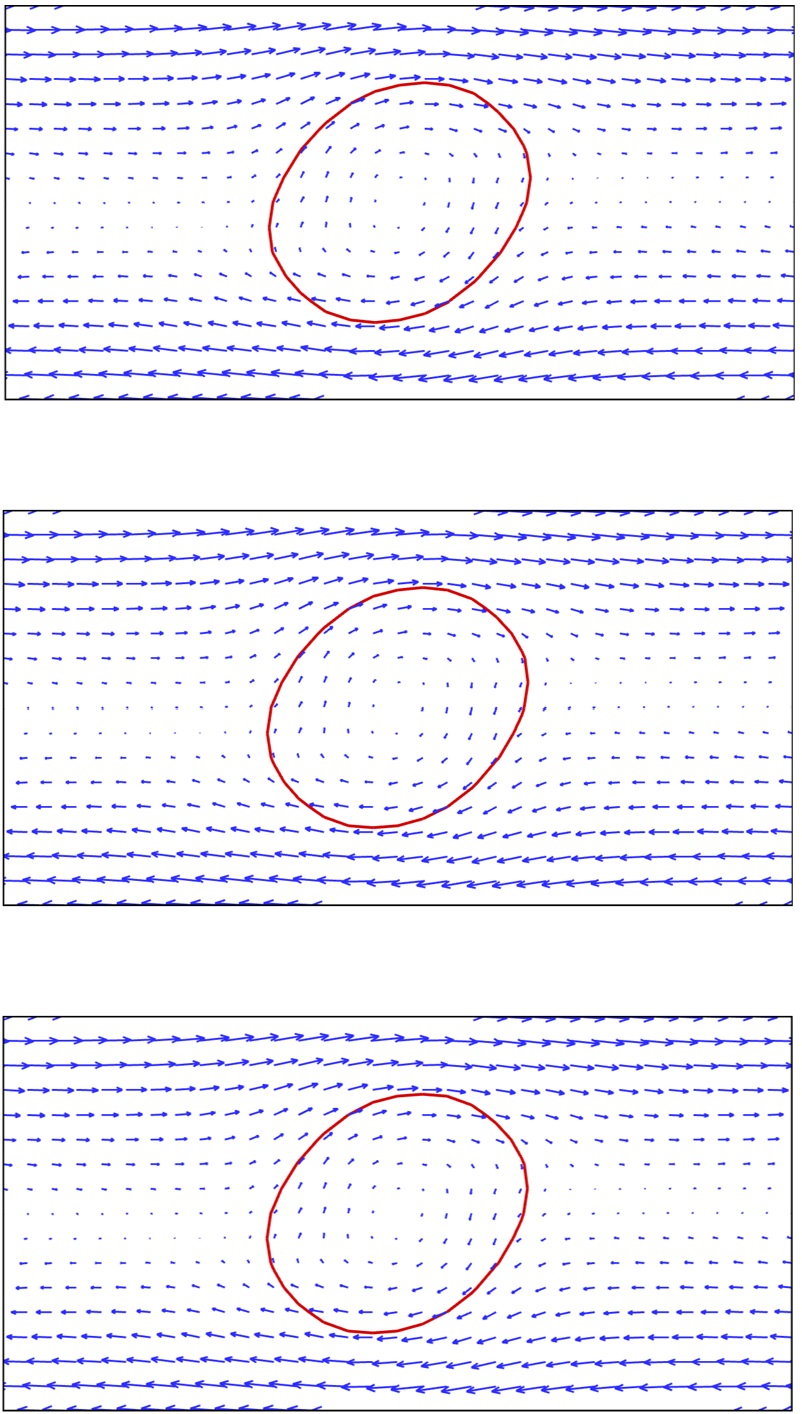}
		\caption{}
		\label{fig8b} 
	\end{subfigure}
	\caption{Evolution of droplet interfacial profile over time, (a) the LB model in Ref. \cite{Liang2014Phase}, (b) the present model.} 
	\label{fig8}
\end{figure}
 
Fig. \ref{fig8} illustrates the interfacial profile of the droplet and velocity vectors evolution with time under shear flow, comparing results calculated utilizing the LB model in Ref. \cite{Liang2014Phase} and those from the current model, respectively. As referenced in Fig. (\ref{fig8a}), when the LB model in Ref. \cite{Liang2014Phase} is applied, the droplet smaller than the critical radius gradually decreases in size and eventually disappears over time. Conversely, Fig. (\ref{fig8b}) demonstrates that the droplets's size is preserved well by the current modified model. The results further indicate that the classical C-H model's limitation in preserving the volumes of the two phases, while the current modified C-H model shows significant improvement in this aspect. This work shows the enhanced capabilities of the current model in handling droplet evolution, with a particular focus on maintaining droplet volume and interface stability. Such advancements enhance the reliability of phase-field methods in simulating two-phase fluid dynamics.

\section{Conclusion}
In this work, we introduce a modified conservaed phase field model for modeling the two-phase flow. This method preserves the local volume of the two phases by utilizing an improved profile correction model, and combines with a signed distance function from the level-set approach to compute the presssion term. Instead of relying solely on the order parameter, we use the correlation between the signed distance function and the order parameter, which helps to avoid issues with jumps and discontinuities at the interface casued by the nonliner sharpening of the fluxes. This enhancement significantly increases the model's accuracy in capturing the two-phase interface and its evolutionary behavior, particularly in addressing the small droplet problem.

After obtaining the diffusion interface model, the LB model is employed to model the two-phase fluid flow. This method employs two distribution functions to solve the hydrodynamic equations and the modified C-H equation, respectively. The effectiveness of the LB model is assessed through simulations of various scenarios, including two stationary droplets suspended in the gas phase, interfacial deformation within a single vortex, Rayleigh-Plateau instability, and droplet deformation under a shear flow. The simulation results, when contrasted with those from the classical C-H equation, illustrate the reliability and precision of the current model and the LB method in maintaining the local volumes of the two phases, especially in the scenario dominated by surface tension effects. However, we would like to emphasized that the model discussed in this study is currently limited to two-phase fluid systems. In future work, we intend to extend the LB method, which preserves the conservation of volume in each phase, to multiphase flow simulations offering a broader range of solutions for addressing complex fluid dynamics problems. 

\section*{CRediT authorship contribution statement}
{\bf{Fang Xiong}}: Writing – original draft, Visualization, Validation, Software, Methodology, Investigation, Formal analysis, Data curation. {\bf{Lei Wang}}: Review \& editing, Supervision, Conceptualization. {\bf{Xinyue Liu}}: Supervision, Software, Methodology, Resources.

\section*{Declaration of competing interest}
The authors declare that they have no known competing financial interests or personal relationships that could have appeared to influence the work reported in this paper.

\section*{Data availability}
Data will be made available on request.

\section*{Acknowledgements}
This work is financially supported by the National Natural Science Foundation of China (Grant No. 12472297).


\end{document}